\documentclass[conference]{IEEEtran}
\usepackage{amsmath, amssymb, graphicx, algorithm, algorithmic, cite}
\usepackage[english]{babel}

\usepackage{flushend}

\begin{document}

\title{Learning the Wireless V2I Channels Using Deep Neural Networks}

\author{\IEEEauthorblockN{Tian-Hao Li$^1$,
        Muhammad R. A. Khandaker$^1$,
        Faisal Tariq$^2$,
        Kai-Kit Wong$^3$ and
        Risala T. Khan$^4$}
\IEEEauthorblockA{
        $^1$School of Engineering and Physical Sciences, Heriot-Watt University, Edinburgh, United Kingdom\\
        $^2$James Watt School of Engineering, University of Glasgow, United Kingdom\\
        $^3$Department of Electronic and Electrical Engineering, University College London, United Kingdom\\
        $^4$Institute of Information Technology, Jahangirnagar University, Dhaka, Bangladesh\\
Corresponding e-mail: $\rm m.khandaker@hw.ac.uk$}}

\maketitle

\begin{abstract}

For high data rate wireless communication systems, developing an efficient channel estimation approach is extremely vital for channel detection and signal recovery. With the trend of high-mobility wireless communications between vehicles and vehicles-to-infrastructure (V2I), V2I communications pose additional challenges to obtaining real-time channel measurements. Deep learning (DL) techniques, in this context, offer learning ability and optimization capability that can approximate many kinds of functions. In this paper, we develop a DL-based channel prediction method to estimate channel responses for V2I communications. We have demonstrated how fast neural networks can learn V2I channel properties and the changing trend. The network is trained with a series of channel responses and known pilots, which then speculates the next channel response based on the acquired knowledge. The predicted channel is then used to evaluate the system performance.

\end{abstract}


%
\IEEEpeerreviewmaketitle

\section{Introduction}
Since the first generation (1G) wireless communication network entered the market in the 1980s,
the world has been dramatically changed by the development of mobile communication
technologies. It has gone through several evolutions in the past few decades, from 1G to 5G,
and due to the huge potential demand all over the world, the technology will continue to
upgrade rapidly \cite{6g_vision}. While 2G, 3G
and 4G were about connecting people and parts of things, 5G will connect everything and it can provide
unlimited access to anywhere, anytime, anybody and anything \cite{what_will_5g_be}.

In recent years, orthogonal frequency division multiplexing (OFDM) has become a popular choice for fast-speed and high-quality communication systems. 
However, channel modelling and channel estimation are two
major challenges affecting the performance of OFDM systems. In order to estimate
the channel response, pilot-based channel estimation is commonly adopted, in which a training sequence
composed of known data symbols (pilots) is transmitted, and the channel parameters are
initially estimated using the received pilot signals \cite{ch_ofdm}. Minimum mean-square error (MMSE)
and least squares (LS) are two traditional estimation approaches. 

Vehicle-to-Infrastructure (V2I) communication is about the data transmission between
vehicles and infrastructure on the roads. V2I communication system is normally wireless and
two-way: infrastructure like traffic lights can provide information to cars and vice-versa.
This communication system can provide quantitative and real-time information
that can be used for safety, mobility, and environmental benefits. When V2I
communications are widely used, revolution of roadways will occur all over the world. For
example, self-driving vehicles will become reality and it is vital to make successful and safe
autonomous cars. To make this huge change happen, the basic theory and techniques
must now be developed urgently.

To enhance the performance of communication systems and to solve signal processing and
communications problems, deep learning (DL) has recently drawn utmost popularity \cite{jiang2016machine, zhang2019deep, conf_v2x_res, drl_v2v_conf}. A deep
neural network (DNN) is an algorithm with learning ability and optimization capability that
can approximate many kinds of functions. Particularly, problems without any precise numerical model
can now be solved using DL methods \cite{ml_vbrief}. 
In order to leverage the advantages of using a large group of data for
communication performance improvement, several machine learning methods including
supervised, unsupervised and reinforcement learning have been proposed based on the traditional
approaches. The machine learning can be useful in analyzing communication environment
variance, making decisions autonomously, transmission routing, network security, and
system resource management \cite{conf_v2x_res, ml_vbrief}.

Recently, some works have concentrated in the area of channel
estimation using DL methods \cite{dnn_ch_dd, ch_bpnn, pow_dl_ch, dnn_ch_bs_mimo, dnn_ch_mmimo}. The authors in \cite{ch_bpnn, pow_dl_ch} proposed the back propagation (BP) learning algorithm to build a multilayered perceptron (MLP) neural network as an estimator
for OFDM communication channels. In \cite{pow_dl_ch}, a method of implicit channel state information (CSI) estimation and direct recovery of transmission symbols based on deep learning has been proposed. The authors in \cite{dnn_ch_bs_mimo} proposed an approximate information passing network for millimetre-wave massive multiple-input and multiple-output (MIMO) systems based on learned denoising which is a deep learning model that can analyze channel structure and estimate channel from a big training database. In \cite{dnn_ch_mmimo}, the authors developed a CSI feedback model by spreading a new learned CSI perception and
restoration architecture. By learning the spatial features directly and combining the
correlation of samples in time-varying MIMO channels, the system greatly improves the
quality of recovery and the trade-off between the compression ratio (CR) and the recovery
quality \cite{dnn_ch_mmimo}. 

All of the above works \cite{dnn_ch_dd, ch_bpnn, pow_dl_ch, dnn_ch_bs_mimo, dnn_ch_mmimo} have developed DL-based channel estimation techniques in traditional quasi-stationery wireless communication systems. However, with the trend of high-mobility wireless communications between vehicles and
vehicles-to-infrastructure (V2I), many rising problems appear because of the high variances in
communication environments, which are fundamentally different from traditional wireless
communication problems. For the communication with large Doppler-shift and fast-varying
channels, the above methods may not work because in fast time-varying fading environment
channel response is difficult to obtain in real time and the outdated CSI exhibits a significant
negative impact on the performance. The critical point to fit high-mobility networks is to develop future informatized and
intelligent vehicles combined with machine learning to benefit the communication networks \cite{ml_v2x_iot}.

In this paper, we address the adverse effects of imperfect
CSI on V2I communication systems by developing a DL-based channel prediction
approach. Two major challenges facing V2I channel estimation addressed in this paper are \textit{i}) the way to find CSI in real-time with the knowledge of vehicle information including position and velocity, \textit{ii}) the
technique to estimate the fast time-variant channel properly and recover the information with
minimum error. The major contributions of this paper are as follows:
\begin{itemize}
\item Firstly, an OFDM modulation based V2I channel estimation technique is introduced as a baseline to verify effectiveness of the DL-based approach.
\item Secondly, DL-based channel prediction method is proposed to deal with the channel
estimation problem in V2I communication system in high-mobility environments.
\item Thirdly,  the performance of the proposed DL-based approach is compared with
other common algorithms through extensive numerical simulations.
\end{itemize}

\begin{figure}[ht!]
\centering
\includegraphics[width=0.8\linewidth]{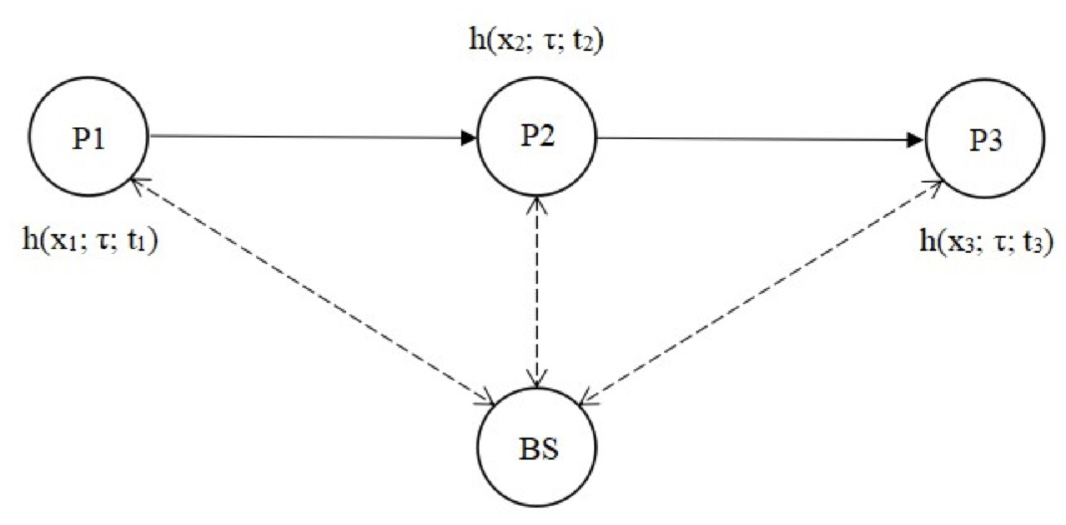}
\caption{The proposed V2V and V2I communication system.} \label{sys_mod}
\end{figure}

\section{System Model}\label{sec_vision}
We consider a high-mobility vehicular communication system (cf. Fig.~\ref{sys_mod}) in which a vehicle intends to communicate with a roadside infrastructure, which can be a traditional cellular base station (BS), traffic lamppost, building or any other fixed structure. At both BS and the vehicle, only one antenna for transmitting and receiving is considered. For simplification, only 2D horizontal plane is considered where the vehicle travels at a constant speed along a straight road. Hence, at the same
position the channel is highly similar even at different time slots, and it can be considered
as position related channel estimation which means that change of channel is related to the
position of vehicle. The channel power gain between the vehicle at position $n$ and the BS is defined as \cite{d2d_v2x}
\begin{align}
    h(n) = G(n)\beta(n)Ad(n)^{-\gamma}, \label{ch_gain}
\end{align}
where $G(n)$ is assumed to be exponentially distributed fast fading power gain, $\beta(n)$ is the log normal shadow fading component, $A$ is the constant pathloss, $d(n)$ is the distance between the vehicle and the BS, and $\gamma$ is the pathloss exponent. For a frequency-selective multipath fading channel, the channel impulse response (CIR) is given by
\begin{align}
h(n) & = \sum_{l=0}^{L-1}C_{l}e^{j\phi_{l}(n)}\delta(n-\tau_l)\\
& = \alpha_l \delta(n-\tau_l),
\end{align}
where $L$ is the total number of multipath component and $\phi_{l}$, $\tau_l$, and $\alpha_l = \sum_{l=0}^{L-1}C_{l}e^{j\phi_{l}(n)}$ are phase, time delay, and the time-dependent complex path gain of the $l$th multipath component. For the V2I channel, the time-variant phase associated with the $l$th path is defined as \cite[Chapter 2]{pmc}
\begin{align}
    \phi_{l}(n) = \phi_{l} - 2\pi c \tau_l / \lambda_{\rm c} + 2\pi f_{{\rm D},l}n, \label{phase}
\end{align}
where $c = f_c \lambda_c$ is the speed of light, $\lambda_c$ is the wavelength of the arriving plane wave (carrier signal), $ \tau_l = d_l/c$, $d_l$ is the length of the $l$th path, and $f_{{\rm D},l}$ is the maximum Doppler shift of the $l$th propagation path.
For OFDM transmission, the $m$th modulated time domain symbols can be expressed as
\begin{multline}
  x_m(n) = \frac{1}{\sqrt{N}}\sum_{l=1}^{N-1}, X_m(l)e^{j2\pi nl/N}, ~ n = 0, \cdots, N-1, \label{x_m}
\end{multline}
where $X_m(l)$ is the $m$th modulated symbol, $n$ is the subcarrier index and $N$ is the total number of subcarriers. Cyclic prefix is then appended to the symbols to prevent inter-symbol interference (ISI). The signals are then transmitted through the wireless channel and the signal received at the BS can be expressed as
\begin{equation}
    y_m(n) = h_m(n)\otimes x_m(n) + w_m(n), \label{y_m}
\end{equation}
where $\otimes$ indicates the convolution operation and $w_m(n)$ is the additive white Gaussian noise for the $m$th subcarrier. At the receiver, the cyclic prefix is removed first from the received signal, followed by parallel conversion to frequency domain by applying fast Fourier transform:
\begin{equation}
    Y_m(k) = \frac{1}{{N}}\sum_{n=1}^{N-1} y_m(n)e^{-j2\pi kn/N}, ~ k = 0, \cdots, N-1. \label{Y_m}
\end{equation}
Thus in the frequency domain, the input-output relationship can be expressed as
\begin{align}
Y(k) = H(k) X(k) + W(k), ~ k = 0, \cdots, N-1.
\end{align}
Consequently, the system can be described as a set of $N$ independent parallel Gaussian channels:
\begin{align}
y_k = h_k x_k + w_k, ~ k = 0, \cdots, N-1. \label{ch_para}
\end{align}
For convenience, we rewrite \eqref{ch_para} using matrix notations as \cite{ch_ofdm}
\begin{align}
{\bf y} = {\bf XFg} + {\bf w}, \label{ch_para}
\end{align}
where ${\bf X}$ is a $N \times N$ diagonal matrix containing ${\bf x} \triangleq \left[x_m(0) ~ x_m(1) \cdots x_m(N-1)\right]^T$ as the main diagonal, ${\bf g} \triangleq \left[g_m(0) ~ g_m(1) \cdots g_m(N-1)\right]^T$,  ${\bf w} \triangleq \left[w_0 ~ w_1 \cdots w_{N-1}\right]^T$ and 
\begin{align}
{\bf F} \triangleq \left[\begin{array}{ccc}
F_{0,0} & \cdots & F_{0,N-1}\\
\vdots & \ddots & \vdots\\
F_{N-1,0} & \cdots & F_{N-1, N-1}
\end{array}\right], \label{mat_F}
\end{align}
is the DFT matrix with $F_{n,k} = \frac{1}{\sqrt{N}} e^{-j2\pi nk/N}$.

The objective of this study is to estimate the CIR $H_m(k)$ from the observation of $Y_m(k)$ with known pilot signal $X_m(k)$. In the following, we will first discuss the conventional channel estimation approaches, and then develop a deep learning based channel estimation algorithm for the V2I system \eqref{Y_m}.

\section{Conventional Channel Estimation Technique}\label{sec_wmmse}
In conventional communication systems, both blind and non-blind channel estimation techniques have been considered for estimating the CIR. Popular channel estimation algorithms include maximum likelihood (ML), least mean square (LMS), minimum mean square error (MMSE) and least square (LS) methods. These methods have been studied thoroughly to estimate CSI within a certain time and frequency range. Although the LS algorithm has a worse performance in time-varying environments compared to the other approaches, its implementation is very simple. On the other hand, the MMSE algorithm can be well-behaved in all general fading
channels with both frequency and time selectivity.

\subsection{The LS Algorithm}
The LS algorithm tries to minimize the squared error (i.e., Euclidian distance) between the transmitted and the received signals which is expressed as
\begin{multline}
    \left\| {\bf y} - {\bf XFg}\right\|^2 =  \left( {\bf y} - {\bf XFg}\right)^H\left( {\bf y} - {\bf XFg}\right). \label{sq_err}
\end{multline}
Taking the derivative of \eqref{sq_err} and equating the derivative to $0$, the LS estimate of the channel frequency response is given by
\begin{align}
\hat{\bf h}_{\rm LS} = {\bf F}\left({\bf F}^H{\bf X}^H{\bf X}{\bf F}\right)^{-1}{\bf F}^H{\bf X}^H{\bf y}. \label{h_ls}
\end{align}
Since the optimal training sequence is orthonormal \cite{jrnl_parafac, apcc11parafac}, \eqref{h_ls} eventually reduces to
\begin{align}
\hat{\bf h}_{\rm LS} = {\bf X}^{-1}{\bf y}.
\end{align}

\subsection{The MMSE Algorithm}
The MMSE channel estimator takes the effect of channel noise into consideration. Meanwhile, the MMSE estimate of ${\bf g}$ is given by 
\begin{align}
\hat{\bf g}_{\rm MMSE} = {\bf R}_{gy}{\bf R}_{yy}^{-1}{\bf y}
\end{align}
where
\begin{align}
{\bf R}_{gy} & = {\rm E}\left\{{\bf gy}^H\right\} = {\bf R}_{gg}{\bf F}^H{\bf X}^H,\\
{\bf R}_{yy} & = {\rm E}\left\{{\bf yy}^H\right\} = {\bf X}{\bf F}{\bf R}_{gg}{\bf F}^H{\bf X}^H + \sigma_w^2{\bf I}_N\\
\end{align}
are the corresponding covariance matrices and $\sigma_w^2$ is the noise variance ${\rm E}\left\{\left|w_k\right|^2\right\}$. Since the columns of ${\bf F}$ in \eqref{mat_F} are orthonormal, the frequency-domain MMSE estimate of ${\bf h}$ is given by \cite{ch_ofdm}
\begin{multline}
\hat{\bf h}_{\rm MMSE} = {\bf F} \hat{\bf g}_{\rm MMSE} =  {\bf F}{\bf R}_{gg}\left[\left({\bf F}^H{\bf X}^H{\bf X}{\bf F}\right)^{-1}\sigma_w^2 + {\bf R}_{gg}\right]^{-1}\\
\times \left({\bf F}^H{\bf X}^H{\bf X}{\bf F}\right)^{-1}{\bf F}^H{\bf X}^H{\bf y}. \label{h_mmse}
\end{multline}

%
%

\section{Proposed Machine Learning Approach}\label{sec_dnn}
It has been shown in \cite{pow_dl_ch} that a multi-layer neural network (MLNN) can provide a very good approximation of channels. Hence in the following, we propose a deep learning based V2I channel estimation algorithm using back propagation technique. 

\begin{figure}[ht!]
\centering
\includegraphics[width=\linewidth]{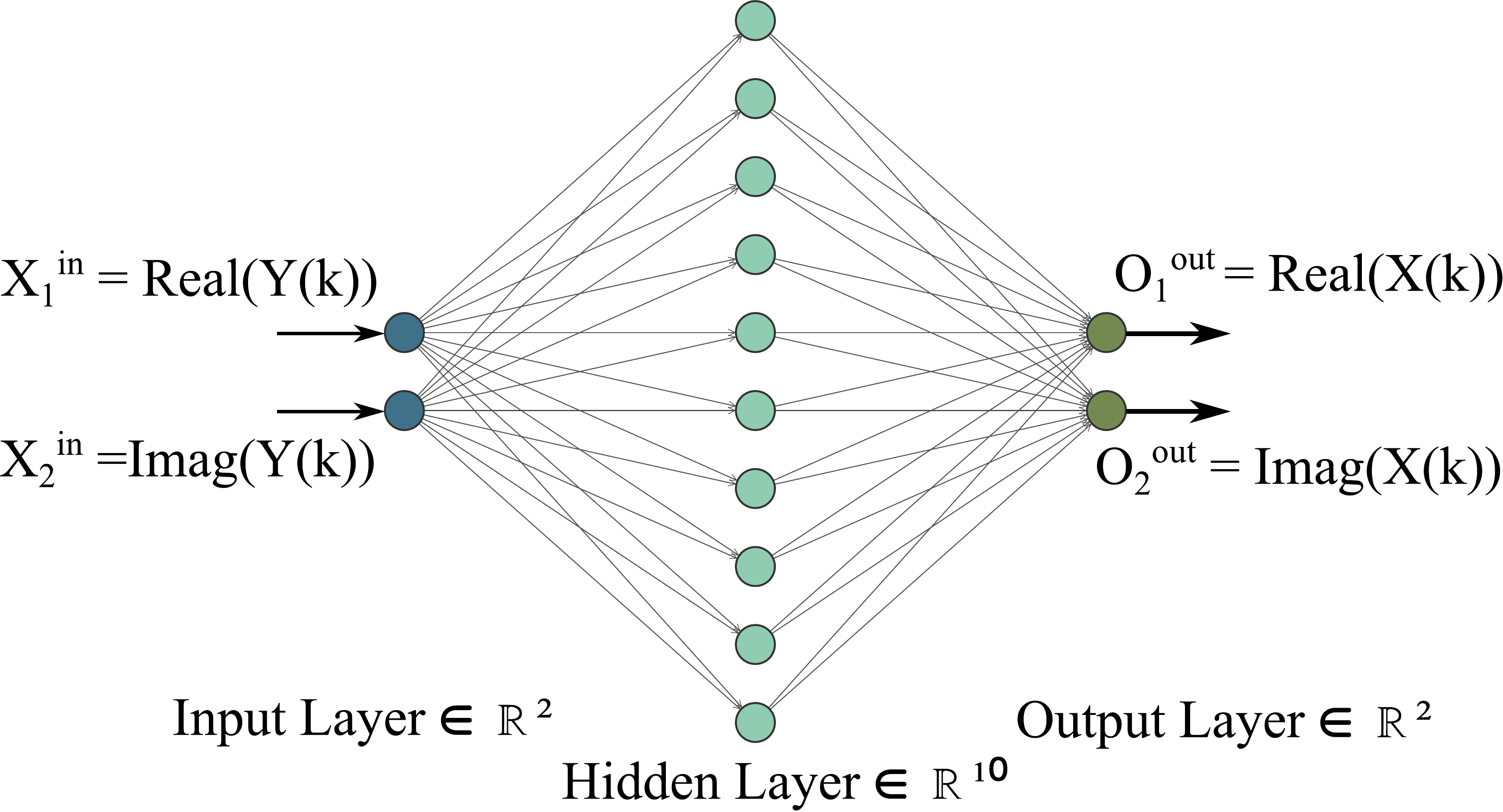}
\caption{The proposed deep neural network for estimating the V2I channel.}\label{fig_dnn}
\end{figure}

\subsection{Defining the Neural Network}
As shown in Fig.~\ref{fig_dnn}, the neural network contains two inputs and two outputs and $n_{\rm h}$ hidden nodes. The inputs to the network are the received signals, and the estimated channel response parameters are the outputs. The two inputs and outputs of the MLNN are connected to the real and imaginary components of the corresponding complex numbers, since neural networks work only with real numbers.
 
 For the machine learning based channel estimation scheme, we consider a supervised learning approach which estimates the CIRs using a fully connected neural network. Note that we do not consider any bias inputs to the neurons at any layer. We apply sigmoid function for the activation of the hidden neurons, while the activation function at the output layer is linear. The output of each neuron is generated by computing the weighted sum of the inputs coming into that node and then applying the activation function. 

Let us now assume that $w_{i,j}^{(1)}$ is the $i$th input to $j$th hidden neuron weight and $w_{j,k}^{(2)}$ is the $j$th hidden node to the $k$th output node weight. The hidden layer activation function is the sigmoid function
 \begin{align}
     \Phi(z) = \frac{1}{1 + e^{-z}} = \frac{e^z}{e^z + 1}. \label{act_hidden}
 \end{align}
 Thus, the hidden layer output is defined as
 \begin{align}
     O_j^{\rm h} = \Phi\left(\text{in}_j\right) = \frac{1}{1 + e^{-\text{in}_j}}, ~ j = 1, \cdots, n_{\rm h}, \label{out_hidden}
 \end{align}
 where $\text{in}_j = \sum_{i=1}^{n_{\rm in}} X_i^{\text{in}} w_{i,j}^{(1)}$, $n_{\rm in}$ is the number of input layer nodes and $X_i^{\text{in}}$ is the $i$th input to the DNN. Similarly, each output node computes its net output as
 \begin{align}
     O_k^{\rm out} = \Phi\left(\text{in}_k\right) = \frac{1}{1 + e^{-\text{in}_k}}, ~ k = 1, 2,  \label{act_out}
 \end{align}
where $\text{in}_k = \sum_{j=1}^{n_{\rm h}} O_j w_{j,k}^{(2)}$. Then, the sum of weighted inputs to the nodes is applied to the output layer activation function which is assumed to be a linear function. Thus the network outputs can be calculated as
 \begin{align}
     O_k^{\rm out} = \Phi\left(\sum_{j=1}^{n_{\rm h}}w_{j,k}^{(2)} \Phi\left(\sum_{i=1}^{n_{\rm in}}  w_{i,j}^{(1)}X_i\right)\right), ~ k = 1, 2.  \label{dnn_out}
 \end{align}

\subsection{Training the Neural Network}
In training process, the weights of all the layers are adjusted according to the mismatches between the outputs and the targets. In each epoch, the direction of changes are that tending to minimize the mean-squared
error (MSE). Accordingly, we define the cost function as the MSE between DNN output and the target output as
 \begin{align}
     \varepsilon = \frac{1}{2}\sum_{k=1}^{n_{\rm o}}\left(t_k - O_k^{\rm out}\right)^2,  \label{cost_fun}
 \end{align}
where $t_k$ is the $k$th desired output and $n_{\rm o}$ is the number of outputs (2 in this case). The gradient descent based back propagation learning rule is exploited to optimize the weights for improving the DNN performance. The gradient descent method minimizes the cost function $\varepsilon$ by updating the weights in the opposite direction of the gradient of the objective function w.r.t. to the weights \cite{grad_desc}. Thus the weight update in each iteration is given by
 \begin{align}
     \nabla w = -\eta \frac{\partial\varepsilon}{\partial w},
 \end{align}
where $0<\eta<1$ is the learning rate that determines the size of the steps in each iteration. Thus the update of hidden-to-output-layer weights can be expressed as
 \begin{align}
     \nabla w_{jk} = \eta \left(t_k - O_k^{\rm out}\right)\frac{\partial O_k^{\rm out}}{\partial w_{jk}} O_j^{\rm h}.
 \end{align}
Note that the gradient of output layer activation function is always unity. Similarly, the back propagated updates of the input-to-hidden-layer weights can be expressed as
 \begin{align}
     \nabla w_{jk} = \eta \left[\sum_{k=1}^{n_{\rm o}}w_{jk}\left(t_k - O_k^{\rm out}\right)\frac{\partial O_k^{\rm out}}{\partial w_{jk}}\right] \frac{\partial O_j^{\rm h}}{\partial w_{ij}}X_i.
 \end{align}

\begin{figure}
\centering
\includegraphics[width=0.8\linewidth]{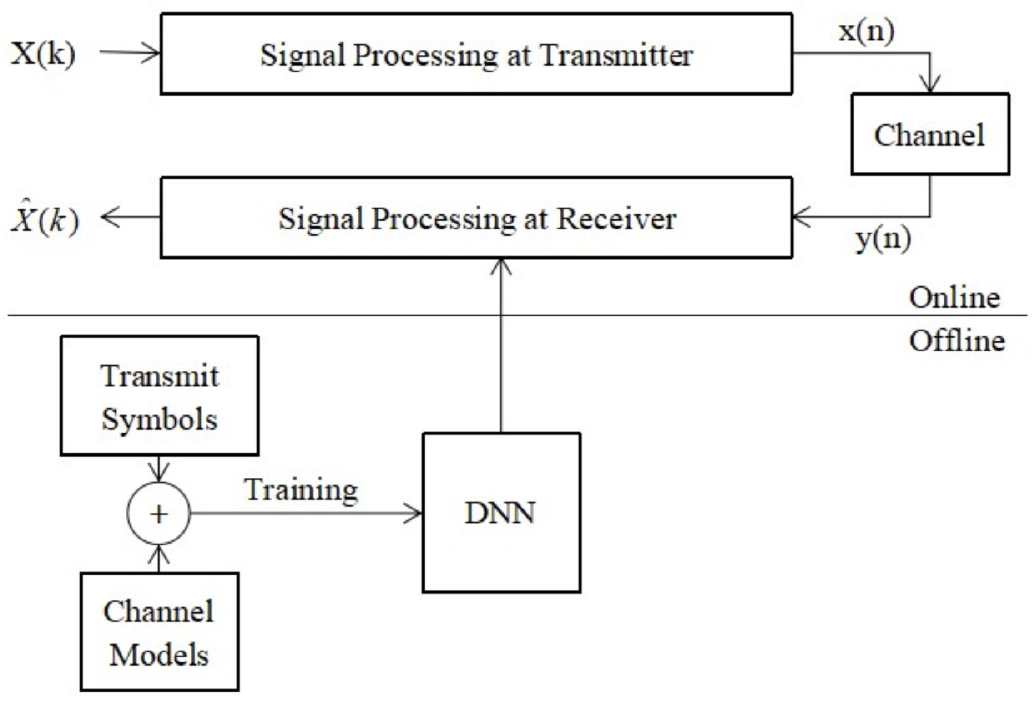}
\caption{Proposed DNN-based channel estimation method.}\label{dnn-ch-estim}
\end{figure}

For training the DNN, we first generate a large set of pilot symbols following a particular modulation scheme. We then use a series of historical channel responses between moving vehicle and BS for optimizing the weights of the neural network. The training process continues until the target accuracy is reached. The overall training procedure is summarized in Fig.~\ref{dnn-ch-estim}.

\section{Numerical Simulations}\label{sec_sim}
In this section, we perform numerical simulations to demonstrate the effectiveness of the proposed machine learning based channel estimation method for wireless V2I communications. Throughout this section, we compare the performance of the proposed approach against the conventional LS and MMSE based estimation schemes in \cite{ch_ofdm}. We first demonstrate the estimation accuracy of the DNN method considering a low-mobility environment. We then test the trained DNN performance for V2I channel estimation.

We assume that the multipath V2I channel has time delays and data is transmitted using 4-QAM modulation. The neural network is designed to have a single hidden layer with $10$ neurons and single output layer with two nodes. Sigmoid and linear activation functions has been utilized in the hidden layer and the output layer, respectively. The DNN algorithm is implemented in two ways. The first one is built from sketch using the structure illustrated in Section~\ref{sec_dnn} (identified as `build' in the figures), and the second one is implemented based on the ANN toolbox in MATLAB, whose performance has also been compared in simulations where applicable.


\begin{figure}
\centering
\includegraphics[width=0.85\linewidth]{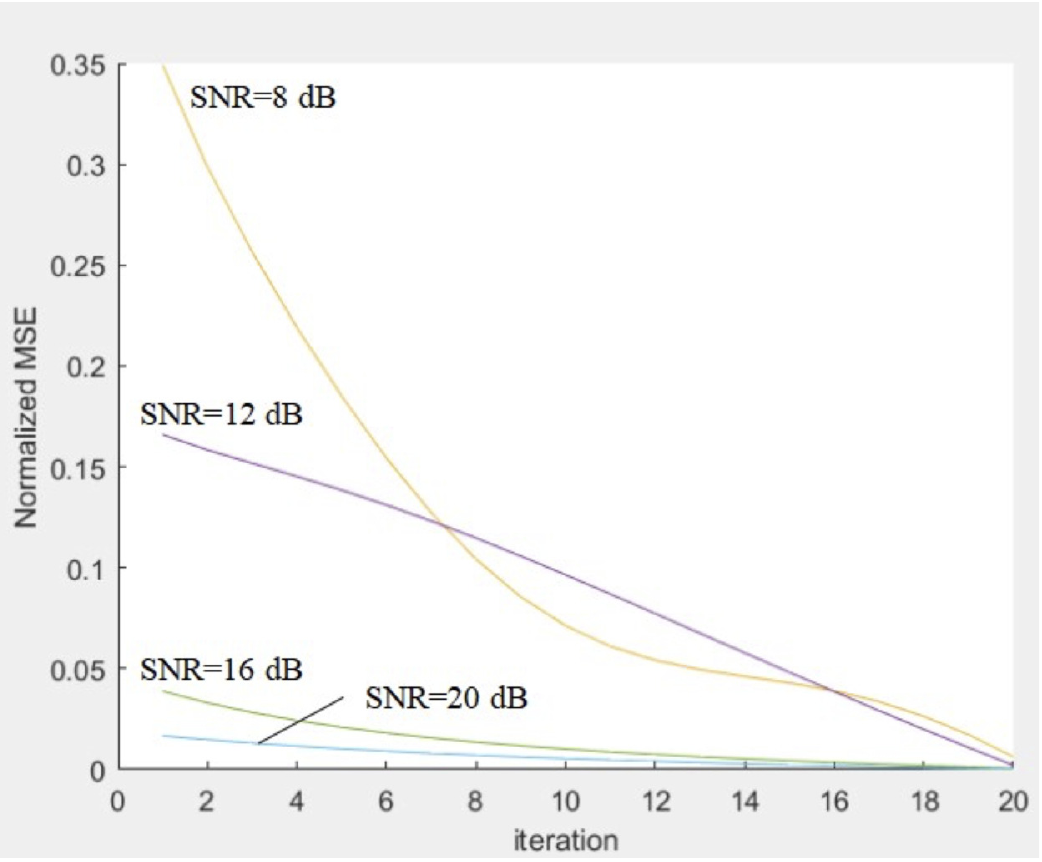}
\caption{MSE performance during the training process.}\label{mse_it}
\end{figure}

\begin{figure}
\centering
\includegraphics[width=0.85\linewidth]{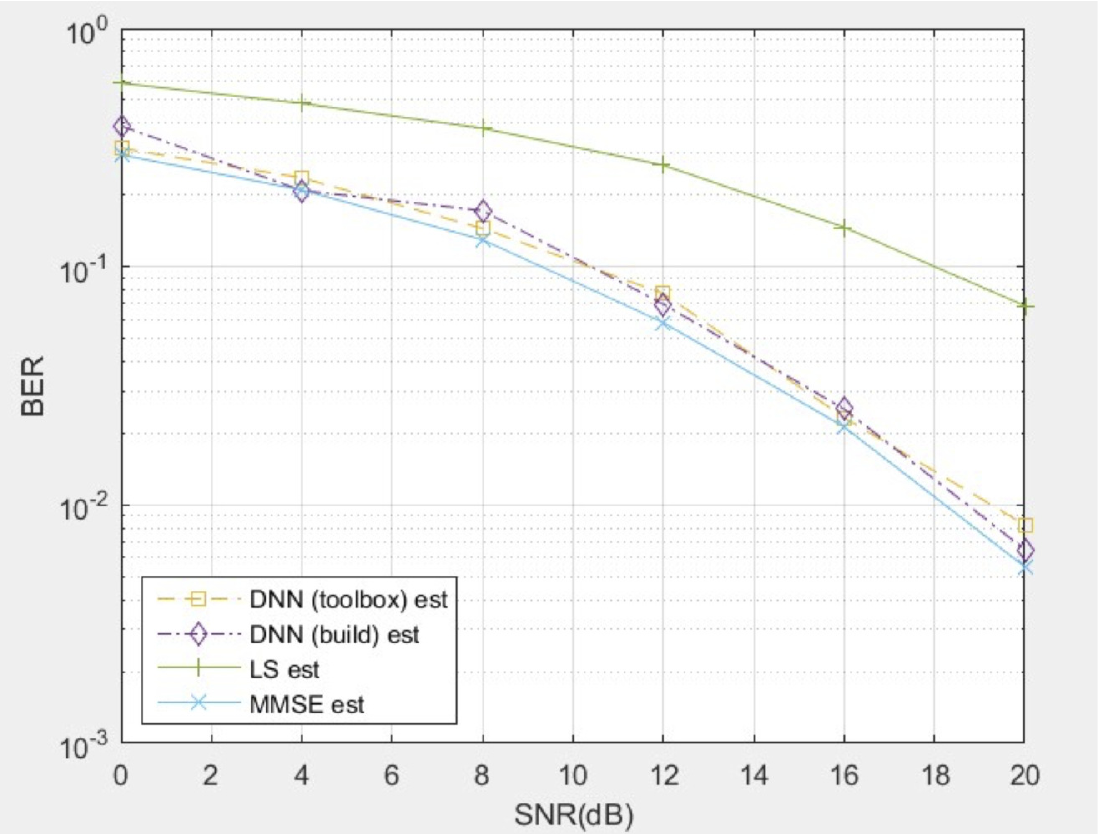}
\caption{BER performance during the training process.}\label{ber_snr}
\end{figure}

Fig.~\ref{mse_it} shows the performance of the proposed DNN approach in the training process according to
normalized MSE at different SNR environment. We run for 20 epochs for each training
data set and It is clear that the error between network outputs and the targets is being
minimized as the training progresses. It can also be found that with higher SNR, which
means the noise power is low compared with signal power, the error of the estimated channel
response becomes lower.

Then, Fig.~\ref{ber_snr} compares the bit error rate (BER) performances of all the methods. It can be seen that the LS algorithm
results in the worst behaviour compared to the others and it cannot correctly estimate the channel
especially when SNR is very low. From this figure it is clear that the performance of proposed back
propagation DNN approaches are much better than the LS algorithm at low SNR values as well as
high SNR values. The results also show that both DNN approaches perform close to each other in a low-mobility environment, and are very close to the MMSE algorithm. Obviously, the MMSE algorithm performs the best according to the simulation results.

\begin{figure}
\centering
\includegraphics[width=0.85\linewidth]{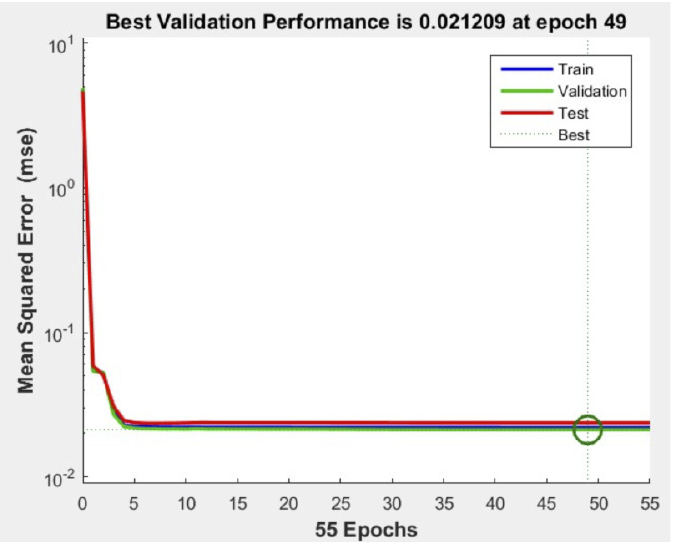}
\caption{MSE at different phases of the DNN appraoch.}\label{mse_epoch}
\end{figure}

\begin{figure}[ht!]
\centering
\includegraphics[width=0.82\linewidth]{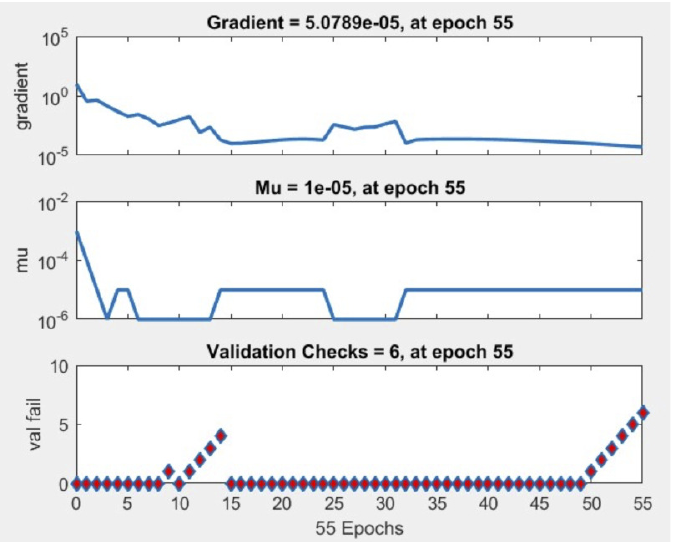}
\caption{Training progression over different epochs.}\label{gd_epoch}
\end{figure}

\begin{figure}
\centering
\includegraphics[width=0.85\linewidth]{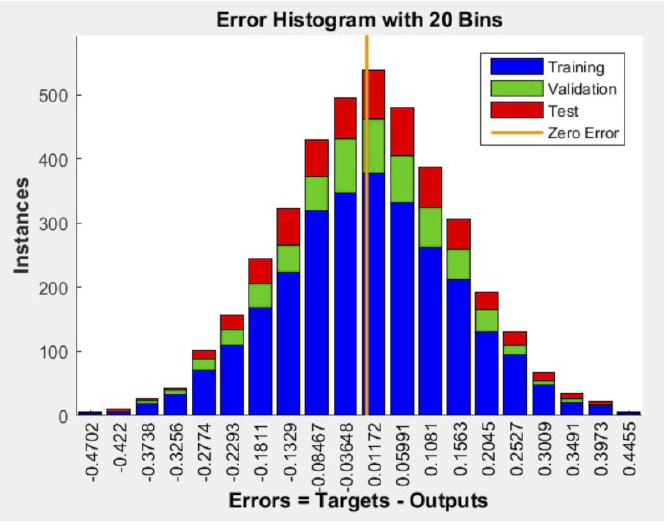}
\caption{Histogram of errors at different processes.}\label{mse_hist}
\end{figure}

\begin{figure}
\centering
\includegraphics[width=0.86\linewidth]{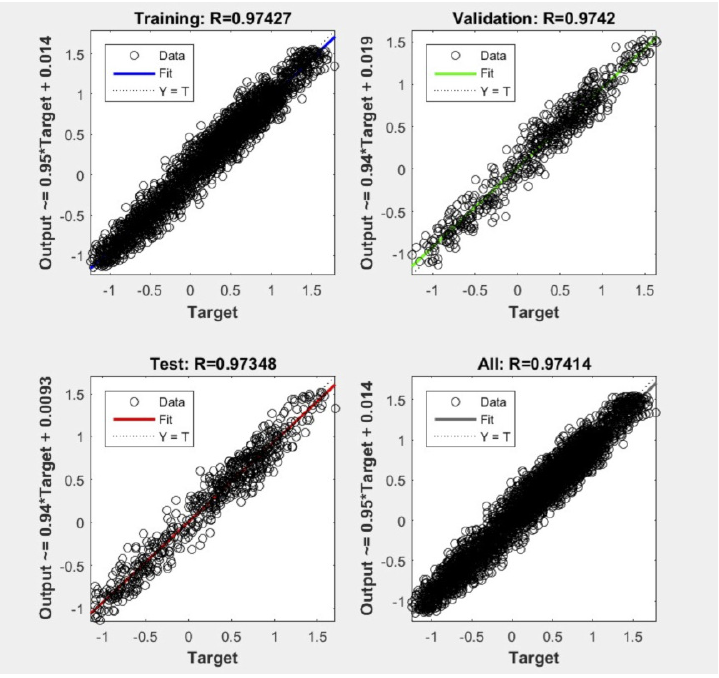}
\caption{Training regression.}\label{regress}
\end{figure}

Next, simulations of a V2I wireless communication system in a fast time-varying fading channel
are carried out to demonstrate the performances of different channel estimation methods. Figs.~\ref{mse_epoch} to \ref{regress} show the training process of DNN for V2I channel prediction. In these simulations, the entire data set is divided into three subsets namely: training (70\%), validation (15\%) and testing (15\%), respectively. During the training process, the MSE reduces sharply and the best estimate is achieved at the 49th epoch with the minimum error distribution. Fig.~\ref{mse_hist} shows the histogram of each subset. From Fig.~\ref{regress}, it can also be seen that the outputs fit the targets very well.

\begin{figure}
\centering
\includegraphics[width=0.86\linewidth]{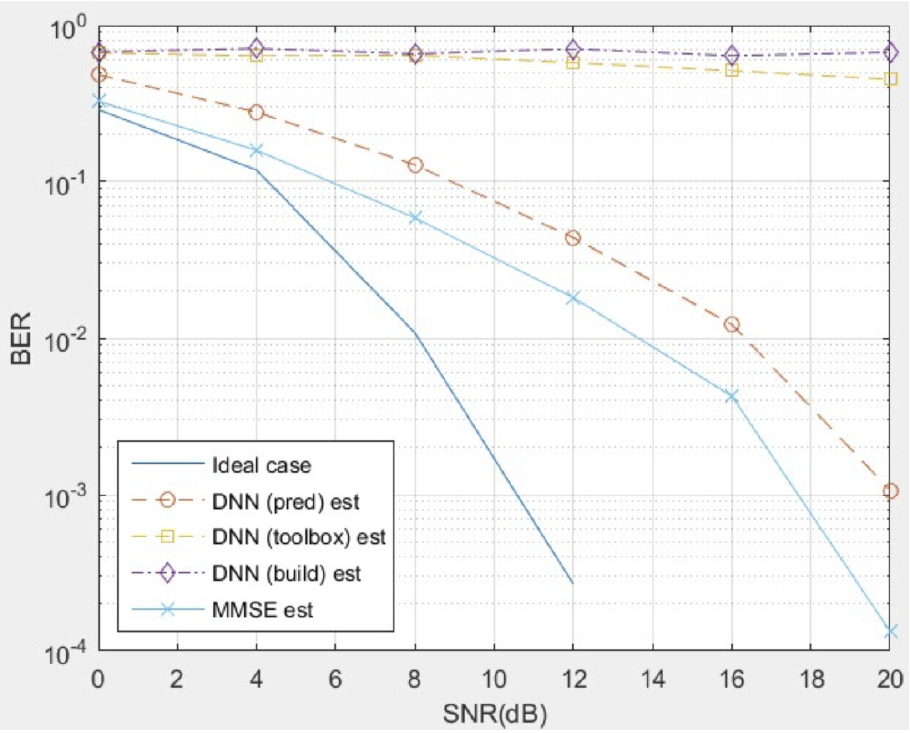}
\caption{BER versus SNR for the testing data.}\label{v2x_ber_snr}
\end{figure}

Finally, we plot the BER performance of the algorithms in Fig.~\ref{v2x_ber_snr}. It can be seen from the results that the BER performance of the traditional DNN methods is significantly worse than the other methods. The reason for this is that the channel estimation in V2I communication
is influenced by many factors, such as fast variation of channels, large Doppler-shift, high
interference and noise. These results further justify the effectiveness of the proposed DL-based approach for V2I channel estimation.


\section{Conclusion}\label{sec_con}
We have introduced a deep learning based channel estimation approach for V2I wireless
communication systems. We have demonstrated that the channels in high-mobility
environment can be estimated using DNN based prediction methods with a group of historical CIR to solve the
outdated CSI problem. Extensive simulation results illustrate that the proposed channel prediction method is able
to dramatically improve the performance of channel estimation in particular in high-mobility environment. Considering the nonuniform movement of vehicles, including variant
position and changing velocity in the training process of DNN can be an interesting future work.

\ifCLASSOPTIONcaptionsoff
  \newpage
\fi

\bibliographystyle{IEEEtran}\footnotesize{

\bibliography{IEEEabrv,refdb}}%

\begin{thebibliography}{10}
\providecommand{\url}[1]{#1}
\csname url@samestyle\endcsname
\providecommand{\newblock}{\relax}
\providecommand{\bibinfo}[2]{#2}
\providecommand{\BIBentrySTDinterwordspacing}{\spaceskip=0pt\relax}
\providecommand{\BIBentryALTinterwordstretchfactor}{4}
\providecommand{\BIBentryALTinterwordspacing}{\spaceskip=\fontdimen2\font plus
\BIBentryALTinterwordstretchfactor\fontdimen3\font minus
  \fontdimen4\font\relax}
\providecommand{\BIBforeignlanguage}[2]{{%
\expandafter\ifx\csname l@#1\endcsname\relax
\typeout{** WARNING: IEEEtran.bst: No hyphenation pattern has been}%
\typeout{** loaded for the language `#1'. Using the pattern for}%
\typeout{** the default language instead.}%
\else
\language=\csname l@#1\endcsname
\fi
#2}}
\providecommand{\BIBdecl}{\relax}
\BIBdecl

\bibitem{6g_vision}
F.~Tariq, M.~R.~A. Khandaker, K.-K. Wong, M.~Imran, M.~Bennis, and M.~rouane
  Debbah, ``A speculative study on {6G},'' \emph{IEEE Commun. Magazine}, June
  2019 (submitted). Available: https://arxiv.org/pdf/1902.06700.pdf.

\bibitem{what_will_5g_be}
J.~G.~A. et~al., ``What will {5G} be?'' \emph{IEEE J. Sel. Areas Commun.},
  vol.~32, pp. 1065--1082, June 2014.

\bibitem{ch_ofdm}
J.~J. {van de Beek}, O.~Edfors, M.~Sandell, S.~K. Wilson, and P.~O. Borjesson,
  ``On channel estimation in {OFDM} systems,'' in \emph{Proc. IEEE 45th Veh.
  Technol. Conf.}, vol.~2, Chicago, IL, USA, 1995, pp. 815--819.

\bibitem{jiang2016machine}
C.~Jiang, H.~Zhang, Y.~Ren, Z.~Han, K.-C. Chen, , and L.~Hanzo, ``Machine
  learning paradigms for next-generation wireless networks,'' \emph{IEEE
  Wireless Commun.}, vol.~24, pp. 98--105, 2016.

\bibitem{zhang2019deep}
C.~Zhang, P.~Patras, and H.~Haddadi, ``Deep learning in mobile and wireless
  networking: {A} survey,'' \emph{IEEE Commun. Surveys \& Tutorials}, 2019.

\bibitem{conf_v2x_res}
J.~Gao, M.~R.~A. Khandaker, F.~Tariq, K.-K. Wong, and R.~T. Khan, ``Deep neural
  network based resource allocation for {V2X} communications,'' in \emph{IEEE
  90th Vehicular Technology Conference: VTC2019-Fall}, 22-25 Sep. 2019,
  Honolulu, Hawaii, USA (submitted). Preprint arXiv:1906.10194.

\bibitem{drl_v2v_conf}
H.~Ye and G.~Y. Li, ``Deep reinforcement learning for resource allocation in
  {V2V} communications,'' in \emph{Proc. IEEE Int. Conf. Commun. (ICC)}, Kansas
  City, MO, 2018.

\bibitem{ml_vbrief}
O.~Simeone, ``A very brief introduction to machine learning with applications
  to communication systems,'' \emph{Available online:
  https://arxiv.org/abs/1808.02342}, 2018.

\bibitem{dnn_ch_dd}
M.~Mehrabi, M.~Mohammadkarimi, M.~Ardakani, and Y.~Jing, ``Decision directed
  channel estimation based on deep neural network k-step predictor for {MIMO}
  communications in {5G},'' \emph{Available: https://arxiv.org/abs/1901.03435},
  Jan. 2019.

\bibitem{ch_bpnn}
N.~Taspinar and M.~N. Seyman, ``Back propagation neural network approach
  forchannel estimation in {OFDM} system,'' in \emph{Proc. Wireless Commun.,
  Netw. Inf. Security (WCNIS)}, June 2010, pp. 265--268.

\bibitem{pow_dl_ch}
H.~Ye, G.~Y. Li, and B.-H. Juang, ``Power of deep learning for channel
  estimation and signal detection in {OFDM} systems,'' \emph{IEEE Wireless
  Commun. Lett.}, vol.~7, pp. 114--117, Feb. 2018.

\bibitem{dnn_ch_bs_mimo}
H.~He, C.~Wen, S.~Jin, and G.~Y. Li, ``Deep learning-based channel estimation
  for beamspace {mmWave} massive {MIMO} systems,'' \emph{IEEE Wireless Commun.
  Lett.}, vol.~7, pp. 852--855, Oct. 2018.

\bibitem{dnn_ch_mmimo}
T.~Wang, C.~Wen, S.~Jin, and G.~Y. Li, ``Deep learning-based {CSI} feedback
  approach for time-varying massive {MIMO} channels,'' \emph{IEEE Wireless
  Commun. Lett.}, vol.~8, pp. 416--419, Apr. 2019.

\bibitem{ml_v2x_iot}
L.~Liang, H.~Ye, and G.~Y. Li, ``Toward intelligent vehicular networks: {A}
  machine learning framework,'' \emph{IEEE IoT Journal}, vol.~6, pp. 124--135,
  Feb. 2019.

\bibitem{d2d_v2x}
L.~Liang, G.~Y. Li, and W.~Xu, ``Resource allocation for {D2D}-enabled
  vehicular communications,'' \emph{IEEE Trans. Commun.}, vol.~65, pp.
  3186--3197, July 2017.

\bibitem{pmc}
G.~L. St\"{u}ber, \emph{Principles of Mobile Communication}, 4th~ed.\hskip 1em
  plus 0.5em minus 0.4em\relax Springer, 2017.

\bibitem{jrnl_parafac}
Y.~Rong, {M. R. A. Khandaker}, and Y.~Xiang, ``Channel estimation of dual-hop
  {MIMO} relay systems using parallel factor analysis,'' \emph{IEEE Trans.
  Wireless Commun.}, vol.~11, pp. 2224--2233, Jun. 2012.

\bibitem{apcc11parafac}
Y.~Rong and M.~R.~A. Khandaker, ``Channel estimation of dual-hop {MIMO} relay
  system via parallel factor analysis,'' in \emph{Proc. 17th Asia-Pacific Conf.
  Commun. (APCC'2011)}, Sabah, Malaysia, Oct. 2-5, 2011.

\bibitem{grad_desc}
S.~Ruder, ``An overview of gradient descent optimization algorithms,''
  \emph{arXiv preprint. Available: arXiv:1609.04747}, June 2017.

\end{thebibliography}

\end{document}